\documentclass[aps,amsmath,amssymb,nofootinbib,superscriptaddress,showpacs,floatfix,prb,twocolumn]{revtex4-1}

\usepackage{latexsym}
\usepackage{graphicx}
\usepackage{times,psfrag,subfigure}
\usepackage{amsmath}
\usepackage{dcolumn}
\usepackage{bm,bbm}       
\usepackage{color}
\usepackage{latexsym,amsmath,amssymb,bm,euscript}
\usepackage{chngcntr}
\counterwithin*{paragraph}{subsection}

\newcommand{\beq}{\begin{equation}}
\newcommand{\eeq}{\end{equation}}
\newcommand{\beqarray}{\begin{eqnarray}}
\newcommand{\eeqarray}{\end{eqnarray}}

\begin{document}

\allowdisplaybreaks

\title{Theory of quasiparticle scattering interference on the surface of topological superconductors}

\date{\today}

\author{Johannes S. Hofmann}

\affiliation{Max-Planck-Institut f\"ur Festk\"orperforschung,
  Heisenbergstrasse 1, D-70569 Stuttgart, Germany} 
\affiliation{Institut f\"ur Theoretische Physik und Astrophysik,
Universit\"at W\"urzburg, Am Hubland, D-97074 W\"urzburg, Germany} 

\author{Raquel Queiroz}
\affiliation{Max-Planck-Institut f\"ur Festk\"orperforschung,
  Heisenbergstrasse 1, D-70569 Stuttgart, Germany}

\author{Andreas P. Schnyder}
\email{a.schnyder@fkf.mpg.de}
\affiliation{Max-Planck-Institut f\"ur Festk\"orperforschung,
  Heisenbergstrasse 1, D-70569 Stuttgart, Germany}

\begin{abstract}
Topological superconductors, such as noncentrosymmetric superconductors with strong spin-orbit coupling, exhibit
protected zero-energy surface states, which possess an intricate helical spin structure. 
We show that this nontrival spin character of the surface states can be tested experimentally
from the absence of certain backscattering processes in Fourier-transform scanning tunneling measurements.  
A detailed theoretical analysis is given of the quasiparticle scattering interference on the surface of 
both nodal and fully gapped topological superconductors  with different crystal point-group symmetries.
We determine the universal features in the interference patterns resulting  from magnetic and nonmagnetic scattering processes of the surface quasiparticles. 
It is shown that Fourier-transform scanning tunneling spectroscopy allows us to uniquely distinguish among different types of topological surface states, such as zero-energy flat bands, arc surface states, and helical Majorana modes, which in turn provides valuable information on the spin and
orbital pairing symmetry of the bulk superconducting state.

\end{abstract}

\date{\today}


% PACS-No:
% 03.65.vf: Topological phases (quantum mechanics)
% 74.50.+r:  tunneling phenomena (superconductivity)
% 74.20.Rp: Pairing symmetries (superconductivity)
% 74.25.F-: Superconductors: transport properties
% 73.20.-r: Surface states, 
% 85.75.-d: spin polarized transport devices (magnetic devices)

\maketitle

\section{Introduction}

An important class of materials for topological superconductivity are noncentrosymmetric superconductors (NCSs).\cite{bauerSigristbook} 
Both fully gapped and nodal NCSs can exhibit nontrivial topological characteristics
manifested by surface states in the form of helical Majorana modes, arc surface states, or zero-energy surface flat bands.~\cite{iniotakisPRB07,satoFujimoto2009,Schnyder2010,schnyderPRB08,qiZhang2010,satoPRL2010,brydon2010,satoPRB2011,tanakaPRL2010,yadaPRB2011,Schnyder2012,tanakaJPSJ2012,dahlhaus2012,matsuura2012} The stability of each of these three types of surface states is ensured by the conservation
of a different bulk topological invariant.
 Interestingly, due to strong spin-orbit interactions,  topological surface states of NCSs
are strongly spin polarized and
possess a highly unusual helical spin texture, where the spin direction varies as a function of surface momentum.\cite{vorontsovPRL08,tanakaNagaosaPRB09,luYip10,brydonNJP2013,2013arXiv1302.3461S}

Due to time-reversal invariance,  surface quasiparticles with opposite momenta have opposite spin polarizations. This in turn leads to the absence of elastic backscattering from nonmagnetic impurities, since scattering processes  
involving  spin flips are forbidden unless  time-reversal symmetry is broken. The absence of  backscattering can be tested experimentally
using  Fourier-transform scanning tunneling spectroscopy (FT-STS).\cite{crommieNat93,capriottiPRB03,wangLeePRB03,franzPRB10,thalmeier2013}  This experimental technique uses the presence of dilute impurities  to probe the electronic properties of surface quasiparticles  at finite momenta ${\bf q}$ through the analysis of interference patterns formed by impurity scattering processes.

\begin{figure*}[t!]
\centering
\begin{minipage}{.69\textwidth}
\includegraphics[clip,angle=0,width=0.9999\columnwidth]{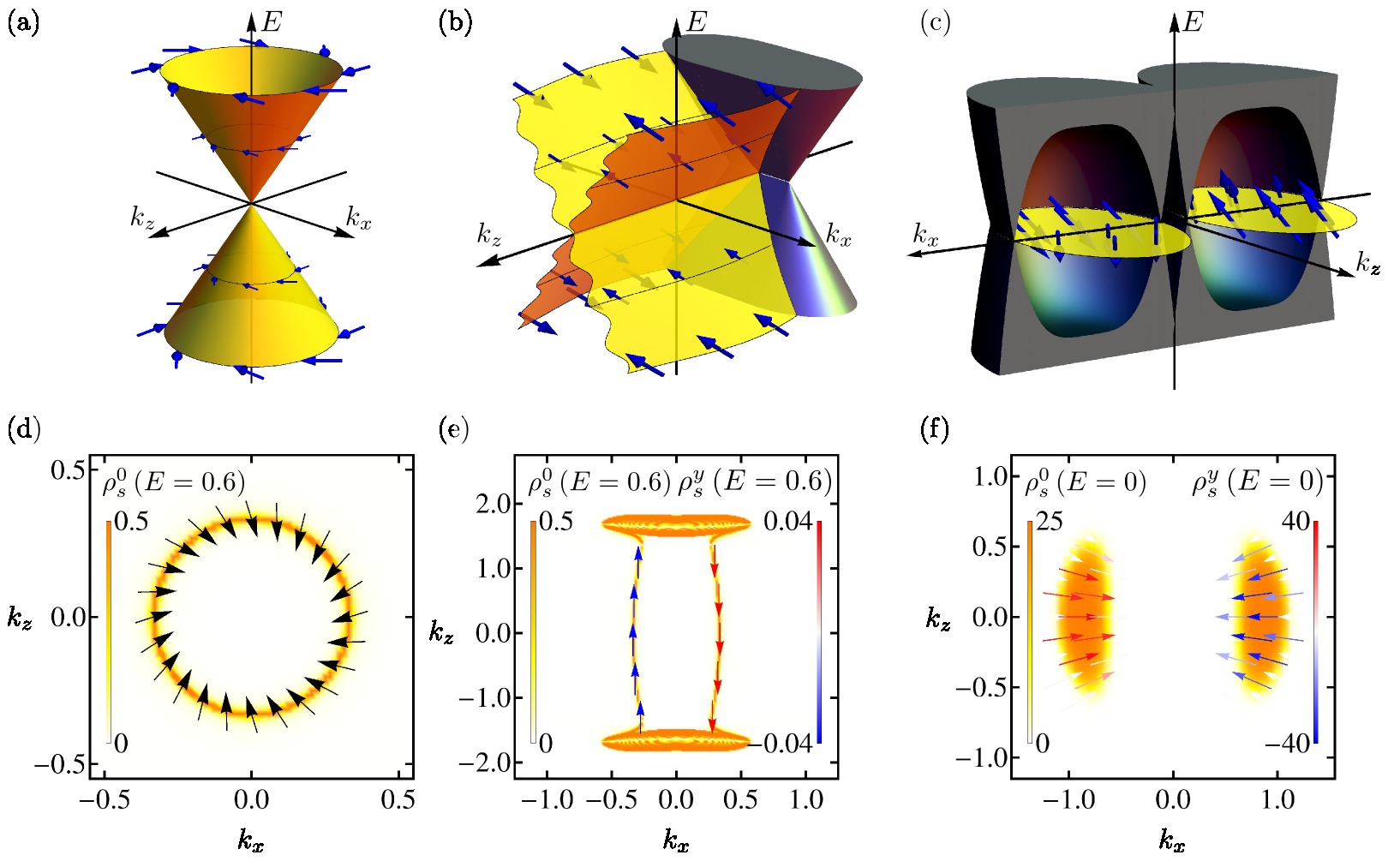}
    \end{minipage}  
    \hfill
    \begin{minipage}[]{.275\textwidth}
\caption{\label{fig1L}  
(Color online) 
Energy spectrum and spin texture $\rho^{\alpha}_{\textrm{s}} (E, {\bf k}_{\parallel} )$
of   the helical Majorana mode [(a) and (d)] and the arc state [(b) and (e)] on the
(010) surface of a topological superconductor with point groups $O$ and $C_{4v}$, respectively [cf.~Eq.~\eqref{modelDef}].
Here we set $\lambda = -2.0$ and $\Delta_{\textrm{s}} =0.5$.
The arrows on the surface states  indicated the magnitude and direction of the spin polarization. 
(c) and (f): Spin texture \mbox{$\rho^{\alpha}_{\textrm{s}} (E=0, {\bf k}_{\parallel} )$} of the zero-energy flat bands on the surface of a nodal NCS
with point group $C_2$, $\lambda = 0.5$, and $\Delta_{\textrm{s}} =4.0$. In-plane and out-of-plane components of 
the spin polarization are represented in (f)  by arrows and by the color scale, respectively.
}  
    \end{minipage}
\end{figure*}

In this paper, we present  an analytical derivation and numerical simulations of the quasiparticle  interference (QPI) on the surface of 
both nodal and fully gapped topological superconductors. For concreteness, we focus 
on single-band centrosymmetric and  noncentrosymmetric superconductors, although our results can be generalized in a straightforward manner to any time-reversal invariant topological superconductor, e.g., to multiband superconductors with dominant spin-triplet pairing.
The surface states of fully gapped topological superconductors are dispersing helical Majorana
modes,\cite{iniotakisPRB07,schnyderPRB08,vorontsovPRL08,tanakaNagaosaPRB09,satoFujimoto2009,luYip10,Schnyder2010,brydonNJP2013,2013arXiv1302.3461S} whereas nodal  topological  superconductors without a center of inversion exhibit  zero-energy surface
flat bands,\cite{tanakaPRL2010,Schnyder2010,brydon2010,satoPRB2011,yadaPRB2011,Schnyder2012} and depending on the crystal point-group symmetry, may also support
zero-energy arc surface states,~\cite{vorontsovPRL08,brydon2010,Schnyder2012} see Fig.~\ref{fig1L}. 
We study the QPI patterns for these three  types of topological surface states in the presence of magnetic and nonmagnetic impurities, and
 identify the universal features in the ordinary and spin-resolved FT-STS response that distinguish among the three types of  surface states.

Interestingly, for helical Majorana modes and arc surface states, we find that the
ordinary QPI patterns resulting from  nonmagnetic impurities
are weak and nonsingular, which is in line with the expected absence of elastic backscattering. 
Similarly, in the case of the flat-band surface states, the absence of backscattering
suppresses the non-spin-resolved FT-STS signal produced by 
nonmagentic scattering processes connecting states with opposite momentum signs.
Magnetic impurities, on the other hand, give rise to a strong and divergent signal in the spin-resolved
FT-STS for all three types of surface states. In the case of the helical Majorana modes, the divergent QPI patterns
exhibit inverse square-root singularities at the momenta $| {\bf q}_{\parallel ,0}|  = 2E / \Delta_{\textrm{t}} $, whereas
for the arc surface states, the divergences in the  FT-STS response at $| q_{x,0} | =  2 E / \Delta_{\textrm{t}}$ 
show a $1 / q_x$ dependence.

The remainder of the paper is organized as follows. We begin in Sec.~\ref{sec:NCS} with
a phenomenological description of time-reversal invariant topological superconductors and their surface states. 
This is followed in Sec.~\ref{sec:QPItop} by an analytical derivation and numerical  simulations of  
  the QPI patterns.
Our summary and conclusions are presented  in Sec.~\ref{sec:Summary}.
Some technical details are given in the Appendices.

%%%%%%%%%%%%%%%%%%%%%%%%%%%%
%%%%%%%%%%%%%%%%%%%%%%%%%%%%

\section{NCS and their surface states}
\label{sec:NCS}

Over the past few years, superconductivity has been discovered in many noncentrosymmetric compounds,\cite{bauerSigristbook}
for example, in Li$_{2}$Pd$_{x}$Pt$_{3-x}$B,\cite{yuanPRL06,nishiyama07} Mo$_3$Al$_2$C,\cite{bauerPRB10,karki10} and BiPd,\cite{joshiPRB11,mondalPRB12} and in the heavy-fermion 
systems CePt$_3$Si,\cite{bauerPRL04} CeIrSi$_3$,\cite{Sugitani06} and CeRhSi$_3$.\cite{Kimura05}
The lack of bulk inversion symmetry in these materials generates   strong Rashba-type spin-orbit couplings  (SOCs) and
allows for a mixing of spin-singlet and spin-triplet pairing states. Due to these properties,  many of the noncentrosymmetric
superconductors are likely to exhibit unconventional pairing symmetries with   nontrivial topological characteristics. 

\subsection{Model definition}

We consider a generic phenomenological description of
single-band NCSs given in terms of
the Bogoliubov-de Gennes (BdG) Hamiltonian $\mathcal{H} = \frac{1}{2} \sum_{\bf k}\Psi_{\bf
    k}^{\dagger}H^{\ }_{\bf k}\Psi^{\ }_{\bf k}$,
with
$\Psi_{\bf k} = ( c_{{\bf k} \uparrow},  c
_{{\bf k} \downarrow},  c^\dag_{-{\bf
k} \uparrow} , c^\dag_{-{\bf k} \downarrow}  )^{\mathrm{T}}$ the Nambu spinor and
\begin{subequations} \label{modelDef}
\begin{eqnarray}
\label{TBmodel}
H_{\bf k} = \begin{pmatrix}
\varepsilon_{\bf k} \sigma_0  + \lambda \, {\bf l}_{\bf k} \cdot \bm{\sigma}  &
\Delta_{\bf k} \cr
\Delta^{\dag}_{\bf k} &  
- \varepsilon_{\bf k} \sigma_0 
+  \lambda \,  {\bf l}_{\bf k} \cdot \bm{\sigma}^{\ast} 
\end{pmatrix} .
\end{eqnarray}
Here, $c^{\dag}_{{\bf k}\sigma}$ denotes the electron creation operator with
spin $\sigma$ and momentum ${\bf k}$, and $\bm{\sigma} = ( \sigma_1, \sigma_2, \sigma_3)$ is the vector of Pauli matrices.
The normal part of the Hamiltonian describes a three-dimensional cubic 
lattice with nearest-neighbor hopping $t$ and chemical potential $\mu$,  
$\varepsilon_{\bf k}
= t\, ( \cos k_x + \cos k_y + \cos k_z) - \mu$,
and antisymmetric Rashba-type SOC  $\lambda \, {\bf l}_{\bf k} \cdot \bm{\sigma}$
with SOC strength $\lambda$.
The superconducting order parameter $\Delta_{\bf k}$ contains both even-parity spin-singlet and odd-parity spin-triplet components 
with 
\begin{eqnarray}
\Delta_{\bf k} = \left(  \Delta_{\textrm{s}} \sigma_0 + \Delta_{\textrm{t}} {\bf d}_{\bf k} \cdot
\bm{\sigma} \right) ( i \sigma_2) ,
\end{eqnarray}
\end{subequations}
where $\Delta_{\textrm{s}}$ and $\Delta_{\textrm{t}}$ represent the spin-singlet and spin-triplet pairing amplitudes, respectively.
In the absence of interband pairing, the  superconducting transition temperature is maximized when the spin-triplet pairing vector ${\bf d}_{\bf k}$ is aligned with the SOC vector ${\bf l}_{\bf k}$.~\cite{Frigeri2004} Hence we assume that
${\bf d}_{\bf k} = {\bf l}_{\bf k}$.
Unless stated otherwise, we set $(t, \mu, \lambda, \Delta_{\textrm{s}}, \Delta_{\textrm{t}} ) = (4.0, 8.0, -2.0, 0.5,  2.0)$
for our numerical calculations and study the QPI patterns as a function of different types of SOC potentials. 
Since the spin polarization of the surface states is generic to NCSs  and the absence of
nonmagentic backscattering is a consequence of time-reversal symmetry, different values for
the parameters $(t, \mu, \lambda,  \Delta_{\textrm{s}}, \Delta_{\textrm{t}} )$ do not qualitatively alter our results.

Crystal lattice symmetries constrain the specific form of the SOC vector ${\bf l}_{\bf k}$.\cite{samokhin09}
Within a tight-binding expansion,
we find that the lowest-order symmetry allowed term for the cubic crystallographic point group $O$, relevant for Li$_2$Pd$_x$Pt$_{3-x}$B and Mo$_3$Al$_2$C, is given by
\begin{subequations}
\begin{eqnarray} \label{SOCcubic}
{\bf l}_{\bf k} 
=
\sin k_x  \, \hat{\bf x}
+
\sin k_y \,  \hat{\bf y}
+
\sin k_z \, \hat{\bf z}.
\end{eqnarray}
Similarly, for the tetragonal point group $C_{4v}$, experimentally represented by
CePt$_3$Si, CeIrSi$_3$, and CeRhSi$
_3$, we obtain
\begin{eqnarray} \label{SOCtetragonal}
{\bf l}_{\bf k} 
=
\sin k_y   \, \hat{\bf x}
-
\sin k_x  \,  \hat{\bf y} .
\end{eqnarray}
Finally, we also consider the monoclinic point group $C
_{2}$, relevant for BiPd, in which case the SOC vector
${\bf l}_{\bf k}$ takes the form
\begin{eqnarray} \label{SOCmonoclinic}
{\bf l}_{\bf k} 
=
\left(\sin k_x + \sin k_y \right) \left(   \hat{\bf x} +  \hat{\bf y} \right)
+
\sin k_z \, \hat{\bf z} .
\end{eqnarray}
\end{subequations}

\subsection{Topological surface states}
\label{sec:SurfStates}

We first discuss the dispersion and spin polarization of the subgap states that appear at the surface of a topological
superconductor described by the lattice BdG Hamiltonian~\eqref{modelDef}. As a result of the strong SOC and the nontrivial topology
of the bulk wave functions, these surface states possess an intricate helical spin texture.  
This is shown in Fig.~\ref{fig1L}, which displays the \mbox{spin-,} \mbox{momentum-,} and
energy-resolved surface density of states,  
\begin{eqnarray} \label{surfDosA}
\rho^{\alpha}_{\textrm{s}} (E, {\bf k}_{\parallel} )
=
- \frac{\hbar}{4 \pi} \mathrm{Im}   \sum_{n=1}^{n_0} 
\mathrm{Tr}  \left\{
S^{\alpha} G^{(0)}_{n  n} ( E, {\bf k}_{\parallel} ) \right\} .
\end{eqnarray}
Here, 
$G^{(0)}_{n n} ( E, {\bf k}_{\parallel} ) = \left[ E + i \eta - H_{n  n} ({\bf k}_{\parallel} ) \right]^{-1}$ 
is the zero-temperature Green's function 
of Hamiltonian~\eqref{modelDef} in a slab geometry with surface perpendicular to the $y$ axis, 
$H_{n n^{\prime} } ({\bf k}_{\parallel} ) = \frac{1}{2 \pi} \int d k_y \, e^{i k_y ( n- n^{\prime} ) } H_{\bm{k}}$,
 ${\bf k}_{\parallel} = ( k_x, k_z)$ denotes the surface momentum, and
$S^{\alpha}=\left( S^0,  S^{\mu} \right)$ stands 
for the charge and spin operators in Nambu space with
\begin{eqnarray} \label{DefSpinOp}
S^{\alpha}
=
\left(
\sigma_3 \otimes \sigma_0,
\sigma_3 \otimes  \sigma_1, 
\sigma_0 \otimes \sigma_2,
\sigma_3 \otimes \sigma_3
\right), \; \; \; \; 
\end{eqnarray}
where $\alpha \in \left\{ 0, 1, 2, 3 \right\}$.
Unless indicated otherwise, 
we evaluate in the following
expression~\eqref{surfDosA} for a slab of thickness $N=10^2$ using an intrinsic broadening of $\eta=0.005$.
The sum in Eq.~\eqref{surfDosA} is taken over the first $n_0=10$ layers, which corresponds approximately to
the decay length of the surface states.
It was recently shown that, depending on the pairing symmetry,  fully gapped or nodal 
topological superconductors of the form of Eq.~\eqref{modelDef}  can support three different types of zero-energy surface states.~\cite{Schnyder2012}

\paragraph{Helical Majorana modes.}  
Fully gapped time-reversal invariant topological superconductors with dominant triplet pairing  
exhibit linearly dispersing helical 
Majorana modes~[see Fig.~\ref{fig1L}(a)].\cite{schnyderPRB08,vorontsovPRL08,PhysRevLett.102.187001}
Similar to the surface states of topological insulators,\cite{liuZhangPRB10,shanShenNJP10} the linearly dispersing Majorana modes of topological superconductors exhibit a helical spin texture, with the spin and momentum directions  locked to each other. 
Interestingly, the spin of  the helical Majorana mode is polarized entirely in the surface plane at all energies.

\paragraph{Arc surface states.} Nodal NCSs with $\Delta_{\textrm{t}} >  \Delta_{\textrm{s}}$ and ${\bf l}$-vector given by Eq.~\eqref{SOCtetragonal} support arc surface states,\cite{bauerSigristbook,brydon2010,iniotakisPRB07,vorontsovPRL08,luYip10,tanakaNagaosaPRB09} i.e., zero-energy states forming one-dimensional open arcs in the surface Brillouin zone, connecting the projections of two nodal rings of the bulk gap~[Fig.~\ref{fig1L}(b)]. We find that the arc states show a strong spin polarization in the $yz$ spin-plane, with a 
vanishing component along the $x$ axis.

\paragraph{Zero-energy flat bands.} Zero-energy flat bands generically occur
at the surface of three-dimensional nodal NCSs
 whose triplet pairing component is comparable or larger than the singlet one.\cite{Schnyder2010,brydon2010,Schnyder2012}
 These flat-band surface states appear within regions of the surface Brillouin zone that are bounded by the projections of the bulk nodal lines [Fig.~\ref{fig1L}(c)]. Strong SOC together with the nontrivial bulk topology  lead to
an intricate three-dimensional spin-texture of the flat-band surface states, as indicated by the arrows and color scale in Fig.~\ref{fig1L}(c).

 \section{QPI of topological surface states}
 \label{sec:QPItop}
 
Weak and dilute scattering potentials on the surface of  topological superconductors modulate the local  density of states of the
surface quasiparticles, thereby leading to quasiparticle scattering interference. Measurements of these interference patterns through 
FT-STS  allow the study of the surface-state dispersion and polarization at finite momentum~${\bf q}$. 
In the following, the $T$-matrix formalism in Born approximation is employed to calculate the QPI spectra of  
subgap states on the (010) surface of topological superconductors. We consider both nonmagnetic  and magnetic  impurities described by
\begin{eqnarray} \label{defHimp}
H^{\beta}_{\textrm{imp}}
=
\sum_{{\bf k} \,  {\bf q}    }
\Psi^{\dag}_{\bf k} V^{\beta}_{\bf q}   \Psi^{\phantom{\dag}}_{ {\bf k} + {\bf q} } , 
\quad \textrm{with} \quad V^{\beta}_{\bf q}= v_{\bf q} S^{\beta} , 
\end{eqnarray}
where $V_{\bf q}^{\beta
= 0}$ corresponds to nonmagnetic  disorder, whereas $V_{\bf q}^{\beta
= 1, 2, 3}$ represents isotropic magnetic exchange scattering caused by impurity spins.
For simplicity, we assume that the magnetic impurities are all
fully polarized along the $\beta$ spin axis by a small external magnetic field of strength $H \ll H_{c2}$. With this, the ordinary $(\alpha
= 0)$ and spin-resolved $(\alpha = 1,2,3)$ FT-STS response $\delta \rho^{\alpha \beta}_{\textrm{s}}$, which is to a good approximation
proportional to the Fourier-transformed differential conductance tensor $d I_{\alpha} /  d V$, is given by\cite{capriottiPRB03,wangLeePRB03}
\begin{eqnarray} \label{FTSTSresponse}
\delta \rho^{\alpha \beta}_{\textrm{s}} (E,  {\bf q}_{\parallel} )
=
- \frac{1}{2 \pi i}
\left[
\Lambda^{\alpha \beta}_{\ } ( E, {\bf q}_{\parallel} ) - \left\{  \Lambda^{\alpha \beta}_{\ } ( E, -{\bf q}_{\parallel} ) \right\}^{\ast}
\right], \quad \;
\end{eqnarray} 
with  ${\bf q}_{\parallel} = ( q_x, q_z)$ the momentum transfer along the surface and
\begin{eqnarray} \label{defAij}
\Lambda^{\alpha \beta}_{\ } (E, {\bf q}_{\parallel} )
=
\int \frac{ d^2 k_{\parallel} }{ ( 2 \pi )^2} 
\sum_{n=1}^{n_0}
 \mathrm{Tr}_{\sigma } \left[
S^{\alpha}  \delta G^{\beta}_{nn} ( E, {\bf k}_{\parallel}, {\bf q}_{\parallel} ) 
\right]_{11} . \quad \;
\end{eqnarray}
In Eq.~\eqref{defAij} the subscript ``$11$'' 
represents indices in Nambu particle-hole space, 
$\mathrm{Tr}_{\sigma }$ is the trace in spin space, and $\delta G^{\beta}_{n n}$ denotes the
change in the BdG Green's function $G^{(0)}_{n n}$ due to scattering processes. For weak impurity potentials $V_{\bf q}^{\beta}$, disorder scattering can be treated within the Born approximation
which yields
\begin{eqnarray} \label{disorderGreen}
 \delta G^{\beta}_{nn} ( E, {\bf k}_{\parallel}, {\bf q}_{\parallel} ) 
 =
\sum_{n' n''} G^{(0)}_{nn'} (E, {\bf k}'_{\parallel} )   
V_{n' n''}^{\beta}  G^{(0)}_{n'' n} (E, {\bf k}_{\parallel}   ) , \; \; \;  \; 
\end{eqnarray}
with ${\bf k}^{\prime}_{\parallel}  ={\bf k}_{\parallel}   + {\bf q}_{\parallel}$.
The Fourier-transformed disorder potential
\begin{eqnarray}
V_{nn'}^{\beta}
=
\frac{1}{2 \pi}
\int d q_y e^{ i q_y ( n - n') } V^{\beta}_{\bf q} 
=
v_0 S^{\beta} \sum_{n^{\prime \prime} =1}^{n_0}
\delta_{n, n''} \delta_{n', n''}  \quad
\end{eqnarray}
describes onsite impurities with strength $v_0 = 0.2$, which are assumed to be equally distributed among the
 $n_0=10$ outermost layers of the superconductor.  
 
 %%%%%%%%%%%%%%%%%%%%%%%%%%%%
\begin{figure*}[t!]
\centering
\includegraphics[clip,angle=0,width=1\textwidth]{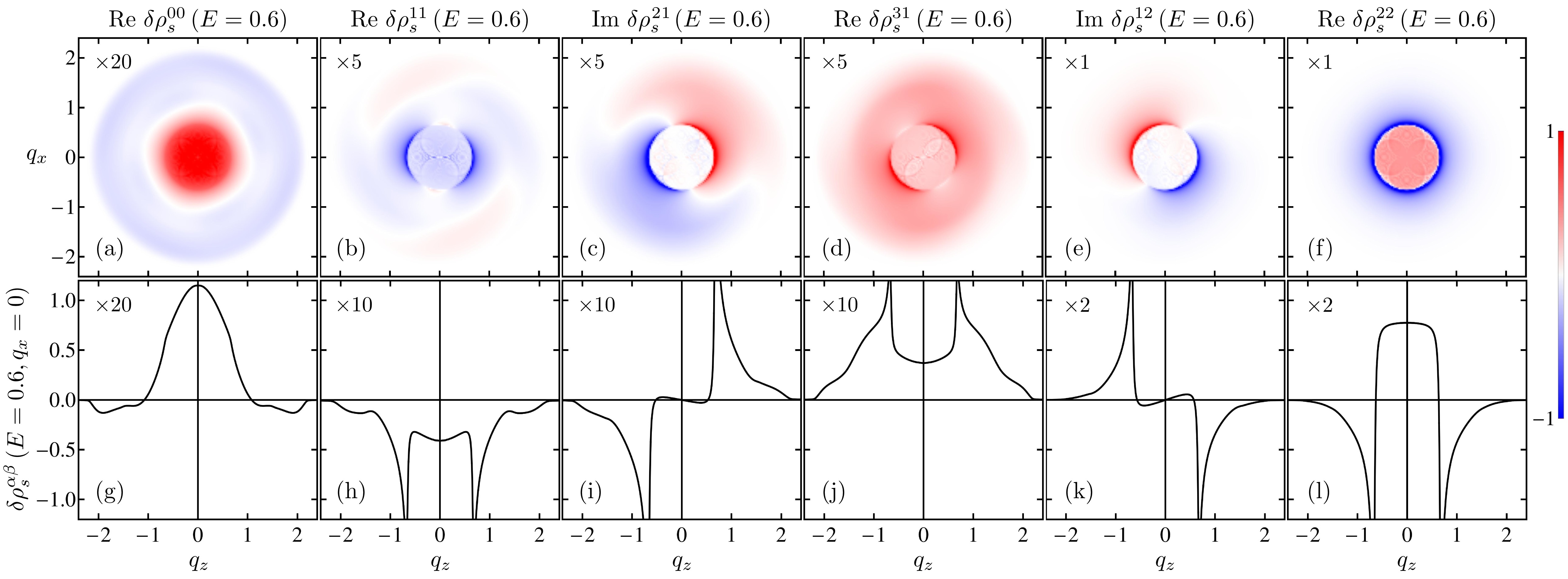}
\caption{\label{fig2L}  
(Color online) Ordinary and spin-resolved FT-STS interference patterns  $\delta \rho^{\alpha \beta}_{\textrm{s}} (E, {\bf q}_{\parallel} )$
arising from magnetic and nonmagnetic scattering processes on the surface
of a fully gapped topological superconductor with cubic point group $O$. The top row shows density plots of $\delta \rho^{\alpha \beta}_{\textrm{s}} (E=0.6, {\bf q}_{\parallel} )$ as a function of  momentum transfer ${\bf q}_{\parallel} = (q_x, q_z)$. The bottom row gives the corresponding cuts
along the $q_x=0$ line. 
Here we set $\lambda=-2.0$ and $\Delta_{\textrm{s}}=0.5$. 
The color scale for the density plots in panels (a)-(d) and the vertical scale in panels (g)-(l) have been multiplied by
a  factor as indicated for clarity.
}  
\end{figure*}
%%%%%%%%%%%%%%%%%%%%%%%%%%%% 

In closing this section, we remark that all the components of the response tensor $\delta \rho^{\alpha \beta}_{\textrm{s}} (E, {\bf q}_{\parallel})$, see Eq.~\eqref{FTSTSresponse}, can, in principle, be measured using spin-polarized
scanning tunneling spectroscopy. That is,  the $\alpha$-spin conductance channel 
can be selected via the polarization direction of the spin-polarized tunneling tip, whereas the component $\beta$ of the spin scattering channel   is determined by the direction of the external magnetic field.

 \subsection{QPI on the surface of centrosymmetric superconductors}

Before discussing numerical simulations of  QPI patterns on the surface of \emph{noncentrosymmetric} superconductors, it is instructive
to first consider \emph{centrosysmmetric} topological systems, whose  quasiparticle surface states can be  described by  simple 
 Dirac-type Hamiltonians. That is, we first study the FT-STS response of helical Majorana modes and 
 arc states on the (010) surface of centrosymmetric superconductors.
 Effective Dirac-type Hamiltonians encoding the low-energy physics of these surface states are derived in Appendix~\ref{appendixA}.
 Due to the simplicity of these surface Hamiltonians, it is possible to derive analytical expressions for the QPI patterns. 
 
 \subsubsection{Helical Majorana modes} \label{majoCentro}
The universal properties of  helical Majorana surface states in superconductors with inversion symmetry are described by the massless Dirac Hamiltonian~[cf.~Eqs.~\eqref{effHammMajo} and \eqref{impProMS}]
\begin{eqnarray} \label{HamWms}
H_{\textrm{MS}} ( {\bf k}_{\parallel} ) 
=
\Delta_{\textrm{t}} \left( k_z \sigma_1 - k_x \sigma_3 \right) + v_0 \sigma_2 ,  
\end{eqnarray} 
where $v_0 \sigma_2$ is an onsite disorder potential describing impurity spins polarized along the $y$ axis. 
Remarkably, as shown in Appendix~\ref{appendixA},  nonmagnetic impurity scattering on the surface of centrosymmetric topological superconductors is forbidden by symmetry, while magnetic impurities only couple via their out-of-plane spin component to the surface states.
As mentioned in Sec.~\ref{sec:SurfStates}, the helical Majorana mode  is spin polarized,
with the spin direction locked to the momentum direction. For the surface state~\eqref{HamWms}, we find that spin and momentum directions are inclined at a right angle to each other. 
This is in contrast to helical Majorana states of \emph{noncentrosymmetric} superconductors, where 
spin and momentum  are in general locked to each other at an angle different from $\pm \pi / 2$, see Fig.~\ref{fig1L}(a) and discussion in Sec.~\ref{sec:Numerics}.

%%%%%%%%%%%%%%%%%%%%%%%%%%%%
\begin{figure*}[t!]
\centering
\includegraphics[clip,angle=0,width=1\textwidth]{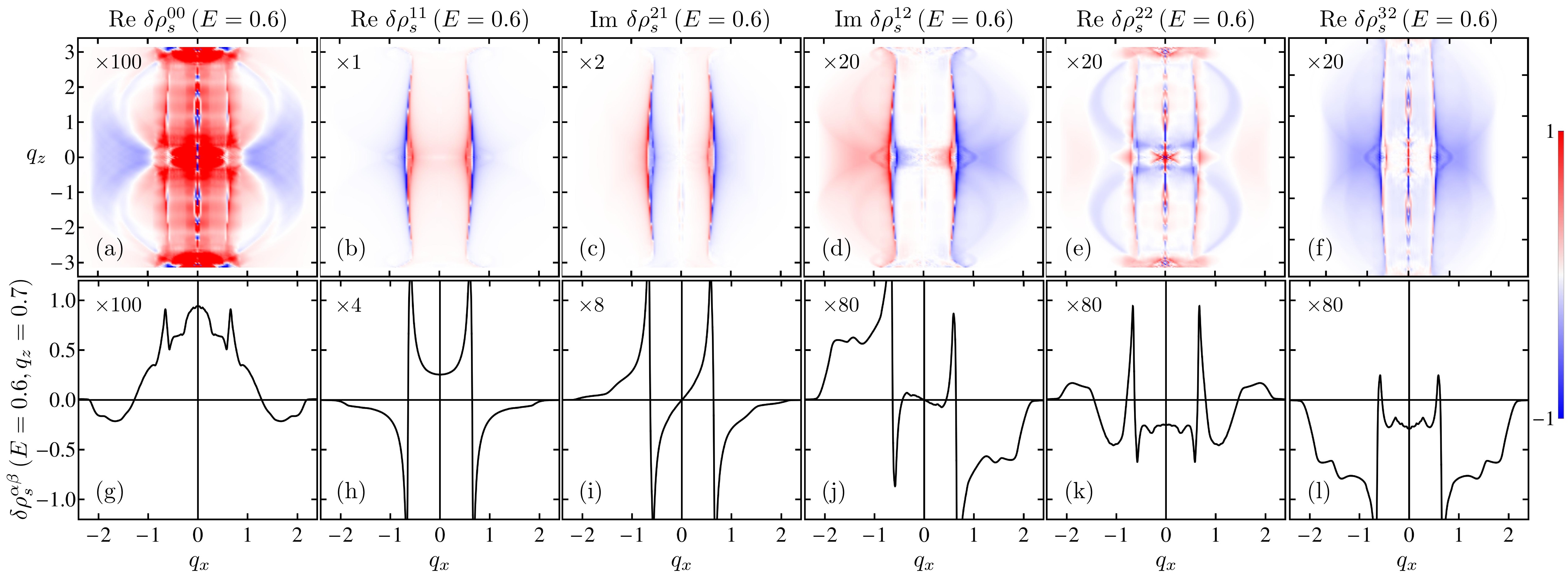}
\caption{\label{fig3L}  
(Color online) Ordinary and spin-resolved FT-STS  response $\delta \rho^{\alpha \beta}_{\textrm{s}} (E, {\bf q}_{\parallel})$ for a nodal 
NCS with tetragonal point group $C_{4v}$, $\lambda=-2.0$, and $\Delta_{\textrm{s}} =0.5$. 
Top and bottom rows show density plots and cuts along the $q_z =0.7$ line, respectively.
The color scale for the density plots in (a)--(f) and the vertical scale in (g)--(l) have been multiplied by a factor as indicated for clarity.
}  
\end{figure*}
%%%%%%%%%%%%%%%%%%%%%%%%%%%%
 
The QPI patterns of the Majorana state~\eqref{HamWms} is obtained from Eq.~\eqref{FTSTSresponse} upon replacing $\Lambda^{\alpha \beta} ( E, {\bf q}_{\parallel})$ by  
\begin{eqnarray} \label{GammaMS}
&&
\Lambda^{\alpha \, \beta=2}_{\textrm{MS}} ( E, {\bf q}_{\parallel} )
=
\\
&& \; \,
\frac{v_0}{2} \int \frac{d^2 k_{\parallel} }{ ( 2 \pi)^2} 
\mathrm{Tr}  \left[ \sigma^{\phantom{\dag}}_{\alpha} G^{(0)}_{\textrm{MS}} ( i \varpi , {\bf k}'_{\parallel} )  \sigma^{\phantom{\dag}}_2  G^{(0)}_{\textrm{MS}} ( i \varpi, {\bf k}_{\parallel}  )  \right]_{i \varpi \to E + i \eta}, 
\nonumber
\end{eqnarray}
where ${\bf k}'_{\parallel} = {\bf k}_{\parallel} + {\bf q}_{\parallel}$, $\varpi$ denotes the Matsubara frequency, and $G^{(0)}_{\textrm{MS}} ( i \varpi , {\bf k}_{\parallel} ) = \left[ i \varpi - \Delta_{\textrm{t}} ( k_z \sigma_1 -k_x \sigma_3 ) \right]^{-1}$ is the bare Green's function of the helical Majorana mode. 
In deriving Eq.~\eqref{GammaMS} we have made use of Eq.~\eqref{SmsOppp} from Appendix~\ref{appMayoEff}. 
Note that since the surface state~\eqref{HamWms} only couples to $y$-polarized impurity spins, the FT-STS signal $\delta \rho^{\alpha \beta}_{\textrm{s}}$ vanishes for $\beta \ne 2$. Inserting the definition of the  Green's function $G^{(0)}_{\textrm{MS}}$
into Eq.~\eqref{GammaMS} we obtain the integral
\begin{equation} \label{MajoIntAB}
\Lambda^{\alpha  2}_{\textrm{MS}}  
=
\int \frac{d^2 k_{\parallel} }{ ( 2 \pi)^2} 
\frac{  (- v_0)  L^{\alpha}_{\textrm{MS}}  }
{ 
\big[ \varpi^2 + \Delta_t^2 k_{\parallel}^2   \big]
\big[ \varpi^2 + \Delta_t^2 \left( {\bf k}_{\parallel} + {\bf q}_{\parallel}  \right)^2  \big]
},
\end{equation}
where the numerator is given by $L^{\alpha}_{\textrm{MS}} =  (0, \Delta_{\textrm{t}} \varpi q_x, \Delta_{\textrm{t}}^2 \, {\bf k}_{\parallel} \cdot {\bf k}'_{\parallel}  + \varpi^2, \Delta_{\textrm{t}} \varpi q_z   )$.
The explicit solution to this integral can be found using the
Feynman parametrization,\cite{peskinSchroeder} 
\begin{align} \label{qpiRespMS}
\Lambda^{\alpha  2}_{\textrm{MS}} ( i \varpi , {\bf q}_{\parallel} )
=
\left\{
\begin{array}{c l}
\frac{ v_0}{ 2 \pi \Delta^2_{\textrm{t}} } \frac{ q_x }{ | {\bf q}_{\parallel} |  }  \zeta \mathcal{F}(\zeta) & \textrm{if $\alpha
=1$} \\
\\
- \frac{ v_0}{ 2 \pi \Delta_{\textrm{t}}^2 } 
\left[
\frac{1}{2}  \ln \left( 1 + \frac{\Omega^2 }{ \varpi^2 } \right) + \mathcal{F}(\zeta)
 \right] 
   & \textrm{if  $\alpha=2$} \\
\\
\frac{ v_0}{ 2 \pi \Delta^2_{\textrm{t}} } \frac{ q_z }{ | {\bf q}_{\parallel} |  } \zeta \mathcal{F}(\zeta) & \textrm{if  $\alpha =3$}, \\
\end{array}
\right. \; \;
\end{align}
and $\Lambda^{\alpha \beta}_{\textrm{MS}}=0$ otherwise. Equation~\eqref{qpiRespMS} is expressed in terms of the
function
\begin{eqnarray}
\mathcal{F} ( \zeta )
=
\frac{1}{ \sqrt{ - \zeta^2 -1 }} \arctan \frac{1}{ \sqrt{ - \zeta^2 -1}} ,
\end{eqnarray}
with the dimensionless variable $\zeta = 2  \varpi / \left( \Delta
_{\textrm{t}}   | {\bf q}_{\parallel} | \right) $ and
$\Omega$ is a momentum cutoff that sets the range of validity
for the Dirac-type Hamiltonian~\eqref{HamWms}. 
 
Interestingly, we find that the ordinary FT-STS signal $\delta \rho^{0 \beta}_{\textrm{s}}$ in the presence of weak magnetic impurities vanishes identically. This is a consequence of time-reversal symmetry, as shown in Appendix~\ref{appendixC}. 
The spin-resolved FT-STS response  $\delta \rho^{\alpha 2}_{\textrm{s}}$, with $\alpha \in \left\{1,2,3 \right\}$, on the other hand, is nonzero and exhibits an inverse square-root singularity at the momenta $ | {\bf q}_{\parallel,0}  | = 2 E / \Delta_{\textrm{t}}$ (cf.\ Fig.~\ref{fig2L}).
This singularity arises due to backscattering processes between states 
at momenta $+{\bf q}_{\parallel, 0}/2$ and $-{\bf q}_{\parallel, 0}/2$.
Furthermore, 
$\delta \rho^{12}_{\textrm{s}}$ and $\delta \rho^{32}_{\textrm{s}}$ have an interesting angular dependence on the momentum transfer ${\bf q}_{\parallel}$
with a twofold symmetry and nodes along the $q_z$ and $q_x$ axes, respectively. 
In contrast, $\delta \rho^{22}_{\textrm{s}}$ is   circularly symmetric in~${\bf q}_{\parallel}$.

 \subsubsection{Arc surface states}
At probe energies $E \ll \Delta_{\textrm{t}}$, the universal physics of the arc surface states
of centrosymmetric superconductors can be captured by the effective low-energy Hamiltonian
 \begin{eqnarray} \label{HeffArc}
H_{\textrm{AS}} ( {\bf k}_{\parallel} ) 
=
- \Delta_{\textrm{t}} k_x \sigma_3 
-
 v_0 \sigma_1 , 
\end{eqnarray} 
where the surface momentum ${\bf k}_{\parallel}=(k_x,k_z)$ is restricted to the range   $\left| k_z \right| < k_z^0$
and $k^0_z$ is half the length of the arc state in the surface Brillouin zone.
 In Eq.~\eqref{HeffArc}
the onsite potential $v_0 \sigma_1$ describes impurity spins polarized along the $x$ axis.
Interestingly,  all other scattering channels are absent due to symmetry constraints.
A detailed derivation of Hamiltonian~\eqref{HeffArc} is presented in Appendix~\ref{appendixAb}.
We observe that the surface state~\eqref{HeffArc} is fully  polarized along the $z$ spin axis, in contrast
to arc states  in nonceontrosymmetric superconductors, which in general show finite spin polarizations
both along the $y$ and $z$ directions, see Fig.~\ref{fig1L}(b).
 
Let us now compute the FT-STS response function for the arc state~\eqref{HeffArc}.
Combining Eq.~\eqref{HeffArc} with Eq.~\eqref{projSaCopAS} we find that the QPI signal is given by Eq.~\eqref{FTSTSresponse}
with $\Lambda^{\alpha \beta} ( E, {\bf q}_{\parallel} )$ replaced by 
\begin{eqnarray}
&&
\Lambda^{\alpha \, \beta=1}_{\textrm{AS}} ( E, {\bf q}_{\parallel} )
=
- \frac{v_0 }{ 2}  (-1)^{\alpha} 
\hspace{-0.15cm}
\int 
\hspace{-0.1cm}
\frac{ d k_\parallel }{ ( 2 \pi)^2}
\\
&& \qquad \quad
\times
\mathrm{Tr}  \left[   \sigma^{\phantom{1}}_{\alpha}   G^{(0)}_{\textrm{AS}} ( i \varpi , {\bf k}^{\prime}_{\parallel}  )  \sigma^{\phantom{1}}_1  G^{(0)}_{\textrm{AS}} ( i \varpi, {\bf k}_{\parallel} )  \right]_{i \varpi \to E + i \eta}, 
\nonumber
\end{eqnarray}
where  
$G^{(0)}_{\textrm{AS}} ( i \varpi, {\bf k}_{\parallel}  ) =  \Theta (k^0_z - | k_z | ) \left[  i \varpi + \Delta_{\textrm{t}}  k_x \sigma_3 \right]^{-1}$
is the  Green's function of the unperturbed system,  $\Theta$ denotes the unit step function, and ${\bf k}'_{\parallel}= {\bf k}_{\parallel} + {\bf q}_{\parallel}$. 
 Because the arc state~\eqref{HeffArc} only interacts with $x$-polarized magnetic
impurities, the QPI pattern $\delta \rho^{\alpha \beta}_{\textrm{s}}$ is identically zero for $\beta \ne 1$.
We now evaluate the above integral by first inserting the bare Green's functions and then performing the
$k_z$ integration. This gives
\begin{equation} \label{LambdaASparII}
\Lambda^{\alpha 1}_{\textrm{AS}} 
=
 \int  \frac{d k_x}{ (2 \pi)^2}
\frac{  v_0  ( \left| q_z \right| - 2 k^0_z )  \Theta ( 2 k^0_z - | q_z | ) L^{\alpha}_{\textrm{AS}}   }
{
\big[ \varpi^2 + \Delta_t^2 k_{x}^2   \big]
\big[ \varpi^2 + \Delta_t^2 ( k_{x} + q_x)^2  \big]
 } ,
\end{equation}
where we have introduced the shorthand notation $L^{\alpha}_{\textrm{AS}} = (0, \Delta_{\textrm{t}}^2 k_x (k_x+q
_x) + \varpi^2, \Delta_{\textrm{t}} \varpi q_x , 0)$.
This integral can be computed explicitly to
\begin{eqnarray} \label{qpiRespAS}
\Lambda^{\alpha 1}_{\textrm{AS}} ( i \varpi, {\bf q}_{\parallel} )
=
\left\{
\begin{array}{c l}
\frac{ v_0 \varpi ( | q
_z | - 2 k^0_z) \Theta ( 2 k^0_z - | q_z | )  }
{ \pi \Delta_t ( 4 \varpi^2 + \Delta^2_{\textrm{t}}  q_x^2 )}   & \textrm{if $\alpha=1$} \\
\\
\frac{v_0 q_{x} 
( | q_z | - 2 k^0_z) \Theta ( 2 k^0_z - | q_z | ) 
}
{ 2 \pi ( 4 \varpi^2 + \Delta^2_{\textrm{t}} q_x^2 )^{\phantom{A}} }   & \textrm{if  $\alpha=2$}, 
\end{array}
\right. \; \;
\end{eqnarray}
and zero otherwise.
 
%%%%%%%%%%%%%%%%%%%%%%%%%%%%
\begin{figure*}[t!]
\centering
\includegraphics[clip,angle=0,width=1\textwidth]{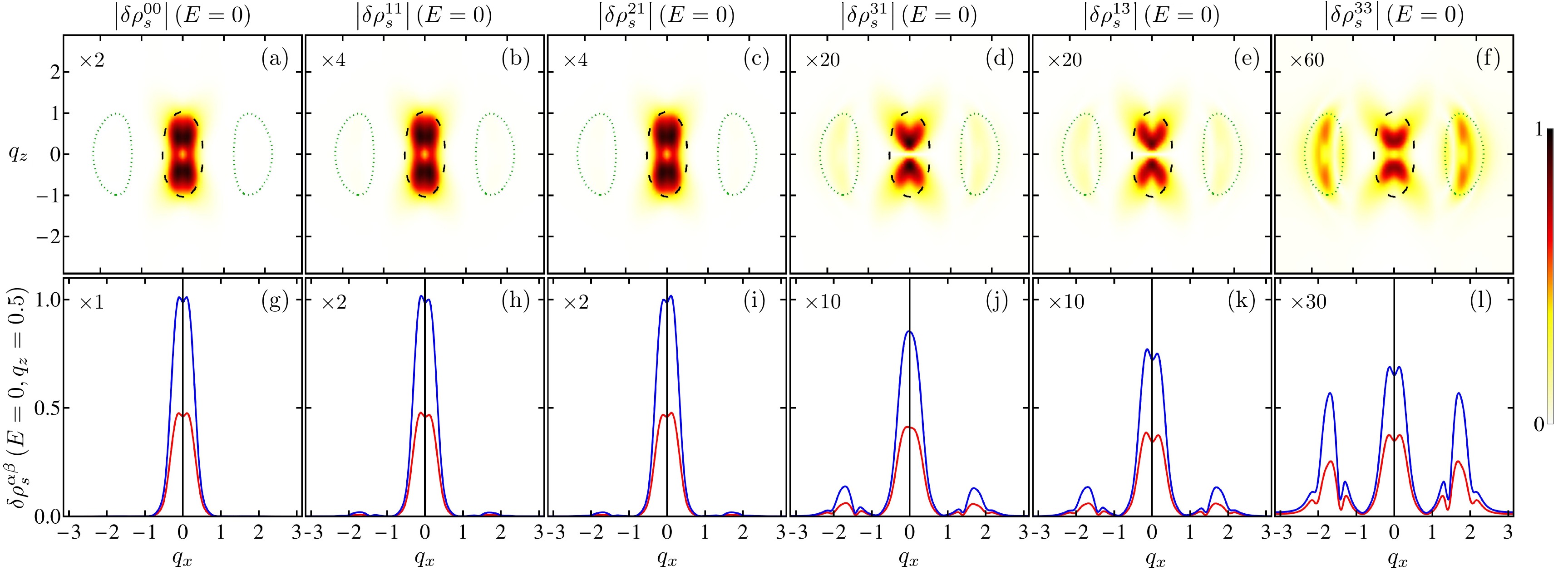}
\caption{\label{fig4L}  
(Color online) Amplitude of the ordinary and spin-resolved FT-STS signal $\delta \rho^{\alpha \beta}_{\textrm{s}} (E, {\bf q}_{\parallel})$ for a nodal 
NCS with flat-band surface states.
Here, the SOC vector ${\bf l}_{\bf k}$ is given by Eq.~\eqref{SOCmonoclinic}, $\lambda=-0.5$, and $\Delta_{\textrm{s}}=4.0$.
The top row shows density plots of $\delta \rho^{\alpha \beta}_{\textrm{s}} (E=0, {\bf q}_{\parallel} )$ as a function of  ${\bf q}_{\parallel}$ for an intrinsic broadening $\eta=0.005$. Green dotted and black dashed lines indicate the boundary of the regions corresponding to
inter-- and intraband scattering, respectively.   
Panels (g)--(l) display  cuts of $\delta \rho^{\alpha \beta}_{\textrm{s}} (E=0, {\bf q}_{\parallel} )$ along the line $q_z=0$ for $\eta = 0.005$ (red) and $\eta = 0.0025$ (blue).
The color scale for the density plots in (b)--(f) and the vertical scale in (h)--(l) have been multiplied by
a  factor as indicated for clarity.
}   
\end{figure*}
%%%%%%%%%%%%%%%%%%%%%%%%%%%%

As before, we find that due to time-reversal invariance the non-spin-resolved QPI patterns resulting from  weak
magnetic impurities are vanishing. The spin-polarized FT-STS signals, however, are finite and show an interesting dependence 
on momentum transfer ${\bf q}_{\parallel}$ (cf.\ Fig.~\ref{fig3L}).
For $\alpha \in \{ 1, 2\}$, $\delta \rho^{\alpha 1}_{\textrm{s}}$ exhibits a $1/q_x$ divergence at $ | q_{x,0} | = 2 E / \Delta_{\textrm{t}}$.
This singularity is due to backscattering processes among states with $x$  momentum component  $+ q_{x,0} /2$ and $-q_{x,0}/2$.
In addition, we find that $\delta \rho^{1 1}_{\textrm{s}}$ is an even function of $q_x$, whereas $\delta \rho^{2 1}_{\textrm{s}}$
is odd in $q_x$.

\subsection{Numerical simulations of QPI patterns on the surface of noncentrosymmetric superconductors}
\label{sec:Numerics}

In this section, we study the FT-STS response of helical Majorana modes, arc states, and zero-energy flat bands on the surface of
\emph{noncentrosymmetric} superconductors. That is, we consider the QPI patterns on the surface of an unconventional superconductor described by Hamiltonian~\eqref{modelDef} with finite SOC strength $\lambda$ and nonzero spin-triplet and spin-singlet pairing components.
Unfortunately, for noncentrosymmetric systems, it is no longer possible to derive 
the FT-STS signal analytically. 
Therefore we resort to 
numerical simulations and compute the QPI patterns through exact diagonalization of the BdG Hamiltonian~\eqref{modelDef}
in a slab geometry with surfaces
perpendicular to the (010) direction.

 \subsubsection{Helical Majorana modes}
 
 In Fig.~\ref{fig2L} is shown the FT-STS response $\delta \rho^{\alpha \beta}_{\textrm{s}}$ of a helical  Majorana mode on the surface of an $O$ point-group NCS  to nonmagnetic
($\beta=0$) and magnetic impurities ($\beta \in \{ 1, 2, 3 \}$). As opposed to centrosymmetric superconductors, we find that Majorana modes of NCSs couple to both nonmagnetic and magnetic scatterers with arbitrary spin polarization (cf.~Sec.~\ref{majoCentro}). Hence, as shown in Appendix~\ref{appendixC}, $\delta \rho_{\textrm{s}}^{0 0}$ and $\delta \rho_{\textrm{s}}^{\alpha \beta}$ with $\alpha, \beta \in \{ 1, 2, 3\}$ are in general nonzero, whereas $\delta \rho_{\textrm{s}}^{0 \beta}$ and $\delta \rho_{\textrm{s}}^{\alpha 0}$ with $\alpha, \beta \in \{ 1, 2, 3\}$ are, due to time-reversal invariance, vanishing within the Born approximation.
Furthermore, we find that four pairs of elements of the FT-STS response tensor are related to each other by crystallographic point-group symmetries (Appendix~\ref{appendixC}). Thus, we plot in Fig.~\ref{fig2L} only the six independent nonzero elements of $\delta \rho_{\textrm{s}}^{\alpha \beta}$, which are either purely real or purely imaginary.

Interestingly, nonmagnetic impurities give rise to only weak, nonsingular FT-STS response, see Figs.~\ref{fig2L}(a) and \ref{fig2L}(g). This is due to the absence of
elastic backscattering processes between states  at momenta $+{\bf q}_{\parallel, 0}/2$ and $-{\bf q}_{\parallel, 0}/2$, where $ | {\bf q}_{\parallel,0}  | =  2 E  / \Delta_{\textrm{t}}$. These nonmagnetic backscattering processes are prohibited by time-reversal symmetry, due
to the opposite spin polarizations of states at opposite momenta. 
 As a result, $\delta \rho^{00}_{\textrm{s}}$ only exhibits a kink at  $ | {\bf q}_{\parallel,0}  |$, but no singularity. 
In the presence of magnetic impurities, however, spin-flip scattering processes  
are allowed. This leads to an inverse square-root singularity in $\delta \rho^{\alpha \beta}_{\textrm{s}}$ at   $ | {\bf q}_{\parallel,0}  | = 2 E / \Delta_{\textrm{t}}$, see Figs.~\ref{fig2L}(h)--(l).

In passing, we point out some interesting features in the angular dependence of $\delta \rho^{\alpha \beta}_{\textrm{s}}$  on transfer momentum ${\bf q}_{\parallel}$. As in the centrosymmetric case,  the dependence of $\delta \rho^{\alpha \beta}_{\textrm{s}}$ on ${\bf q}_{\parallel}$ in Figs.~\ref{fig2L}(b)--(e)  exhibits a $\pi$ rotational symmetry about the origin, i.e., $\delta \rho_{\textrm{s}}^{\alpha \beta} ( {\bf q}_{\parallel} ) 
= [ \delta \rho^{\alpha \beta}_{\textrm{s}} ( - {\bf q}_{\parallel} ) ]^{\ast} $.
Twofold symmetries with high-symmetry lines along the  vertical or horizontal axes, however, are absent. 
We observe that the different angular dependence between the centrosymmetric and noncentrosymmetric cases
is due to differences in the spin polarization [see Fig.~\ref{fig1L}(d)]. While in the centrosymmetric case, spin and momentum of the Majorana
mode~\eqref{qpiRespMS} are locked to each other at a right angle, in NCSs
the angle between spin and momentum directions of the Majorana surface states differs from $\pm \pi /2$ and, moreover, varies strongly
as a function of distance from the surface layer. This dependence of the spin polarization on  layer index $n$ results in the absence
of any twofold mirror symmetries in the QPI patterns of Figs.~\ref{fig2L}(b)--(e).

 \subsubsection{Arc surface states}
 
The FT-STS response of an arc  state on the surface of a $C_{4v}$ point-group NCS  is shown in Fig.~\ref{fig3L}. 
In order to discuss  energy and momentum dependence of these QPI patterns, we first point out that the arc surface state can essentially be
viewed as a quasi-one-dimensional analog of the two-dimensional Majorana mode of the previous subsection. In other words, a description of the arc state can be obtained from the $O$ point-group NCS by interchanging $x$ and $y$ components of the spin operator and by setting $k_z=0$, see Eqs.~\eqref{SOCcubic} and \eqref{SOCtetragonal}. 
This explains the similarities in the QPI patterns of Fig.~\ref{fig3L} with the response at $q_z=0$ shown in Fig.~\ref{fig2L}.

As in Figs.~\ref{fig2L}(a) and \ref{fig2L}(g), we find that the FT-STS signal produced by nonmagnetic impurities is weak and nonsingular, since spin-flip backscattering is prohibited by time-reversal symmetry. Hence $\delta \rho_{\textrm{s}}^{00}$ in Figs.~\ref{fig3L}(a) and \ref{fig3L}(g) only shows a nondivergent kink- or peaklike feature at $ |  q_{x,0} |=  2 E / \Delta_{\textrm{t}}$.
Magnetic impurities, on the other hand, give rise to a strong and divergent response in the spin-resolved
FT-STS [see Figs.~\ref{fig3L}(b)--(f) and \ref{fig3L}(h)--(l)]. Similar to Eq.~\eqref{qpiRespAS}, there is a   divergence 
in $\delta \rho_{\textrm{s}}^{\alpha \beta}$ at $   | q_{x,0} | =  2 E / \Delta_{\textrm{t}}$.
We note that due to the different dimensionality of the momentum phase space, this is not an 
inverse square-root singularity as in Figs.~\ref{fig2L}(h)--(l), but shows a $1/q_x$ dependence.

 \subsubsection{Zero-energy flat bands} 
 
Finally, we discuss the FT-STS response of zero-energy flat bands on the surface of a $C_2$ point-group NCS, which is 
shown in Fig.~\ref{fig4L}.  
As before, we find that due to time-reversal symmetry the only nonzero elements of the response tensor
 $\delta \rho_{\textrm{s}}$ are $\delta \rho^{00}_{\textrm{s}}$ and $\delta \rho^{\alpha \beta}_{\textrm{s}}$ with $\alpha, \beta \in \{ 1, 2, 3 \}$. 
 Since lattice point-group symmetries relate four pairs of  entries of  $\delta \rho_{\textrm{s}}$ to each other (see Appendix~\ref{appendixC}),
 we plot in Fig.~\ref{fig4L} only the six independent nonzero elements of $\delta \rho^{\alpha \beta}_{\textrm{s}}$.
 
The $C_2$ point-group NCS as defined by Eqs.~\eqref{modelDef} and \eqref{SOCmonoclinic} exhibits two different zero-energy flat bands on the (010) surface, one with negative surface momentum $k_x < 0$ and  one with positive momentum $k_x > 0$, see Figs.~\ref{fig1L}(c) and \ref{fig1L}(f).  Hence, in the presence of impurities, it is useful to distinguish between interband scattering with transfer momentum $| q_x | \lesssim 1.0 $ and intraband scattering  with $| q_x | \gtrsim 1.0 $, as indicated in Figs.~\ref{fig4L}(g)--(l) by the green dotted and black dashed ellipses, respectively. 

Due to the opposite spin polarizations of the two zero-energy flat bands,  time-reversal-preserving interband scattering is considerably suppressed.\cite{2013arXiv1302.3461S} 
Hence the part of the ordinary FT-STS signal $\delta \rho^{00}_{\textrm{s}}$ that corresponds to interband scattering [green dotted ellipses in Fig.~\ref{fig4L}(a)] is very weak, whereas the one corresponding to intraband scattering [black dashed ellipses in Fig.~\ref{fig4L}(a)] is strong and divergent. Magnetic impurities, on the other hand, give rise to both strong inter- and intraband backscattering. Consequently, the FT-STS response
shown in Figs.~\ref{fig4L}(b)--(f)  and \ref{fig4L}(h)--(l) exhibits strong divergences both for large  and small transfer momenta, i.e., within the regions in Figs.~\ref{fig4L}(b)--(f) bounded by green dotted and   black dashed lines, respectively.

\section{Summary and discussion}
\label{sec:Summary}

In summary, we have determined the universal features in the QPI patterns produced by magnetic and nonmagnetic impurities on the surface of time-reversal invariant topological superconductors. 
An explicit analytical expression was found for the energy and momentum dependence of the QPI patterns on the surface of centrosymmetric topological superconductors.
For noncentrosymmetric systems, on the other hand, we have numerically determined the ordinary and spin-resolved FT-STS response.

We have studied both fully gapped and nodal topological superconductors and considered three different types of surface states: helical Majorana modes, arc surface states, and zero-energy flat bands. Due to SOC, these surface states exhibit an
 intricate helical spin texture, where the spin polarization strongly depends on the surface momentum.
Time-reversal invariance ensures that surface states with opposite momenta have opposite spin polarizations,
which leads to the absence of backscattering from nonmagnetic impurities.
As a consequence, the ordinary FT-STS signal of Majorana modes and arc surface states due to nonmagnetic scattering
is weak and nondivergent. In the case of the flat-band surface states, the absence of backscattering results in a suppression of the QPI signal
produced by scattering processes with large momentum transfer. In the presence of magnetic impurities, however, spin-flip scattering is possible, and hence backscattering leads to a large and divergent FT-STS response for all three types of surface states. 

 Our results highlight the unique power of the FT-STS technique to detect topological surface states
in unconventional superconductors. We have demonstrated that the FT-STS response allows to
clearly distinguish among the three different types of surface states that  generically occur in
time-reversal invariant topological superconductors. Moreover, the analysis of QPI patterns
can be used to infer information about the pairing symmetry 
and the topological characteristics of the
superconducting state.

\acknowledgments
The authors thank F.~Assaad, P.~Brydon, C.~Timm,  P.~Wahl, and A.~Yazdani  for useful discussions.

%%%%%%%%%%%%%%%%%%%%%%%%%%%%%%%%%%%%%%%%%%%%%%%
%%%%%%%%%%%%%%%%%%%%%%%%%%%%%%%%%%%%%%%%%%%%%%%

\appendix

\section{Low-energy models for the surface states \\ of centrosymmetric superconductors}
\label{appendixA}
In this appendix, we derive low-energy effective Hamiltonians describing  the surface states of
\emph{centrosymmetric}  topological superconductors with time-reversal symmetry. To that end, we consider  
Hamiltonian~\eqref{modelDef} with vanishing SOC strength $\lambda$, zero singlet pairing component $\Delta_{\textrm{s}}$,
and   surface perpendicular to the $y$ axis.
The surface plane is chosen to be at $y=0$, where the bulk superconductor and the vacuum occupy the
half-spaces $y>0$ and $y<0$, respectively. 
The derivation of the surface states of topological centrosymmetric superconductors proceeds along similar lines as for
the case of topological insulators.\cite{liuZhangPRB10,shanShenNJP10}

\subsection{Helical Majorana modes}  
\label{appMayoEff}

First, we examine helical Majorana modes that appear at the surface of
fully gapped topological superconductors.
For concreteness, we consider a centrosymmetric system with cubic
crystallographic point group $O$, i.e., Hamiltonian~\eqref{modelDef} 
with $\lambda=0$, $\Delta_{\textrm{s}}=0$, and
 ${\bf d}$-vector ${\bf d}_{\bf k}$ given by Eq.~\eqref{SOCcubic}. Focusing on low energies, we perform a small-momentum expansion 
near the $\Gamma$ point. This yields
\begin{eqnarray} \label{modDefTildO}
\widetilde{H} ( {\bf k} ) 
=
\begin{pmatrix}
\tilde{\varepsilon}_{\bf k} \sigma_0 & i  \Delta_{\textrm{t}}  \left( {\bf k} \cdot \bm{\sigma} \right)   \sigma_2  \cr
-  i \Delta_{\textrm{t}}    \sigma_2    \left( {\bf k} \cdot \bm{\sigma} \right) &  -  \tilde{\varepsilon}_{\bf k} \sigma_0
\end{pmatrix} ,
\end{eqnarray}
where $\tilde{\varepsilon}_{\bf k} = 3 t - \mu - \frac{t}{2} ( k^2 + k_y^2 )$ and
$k^2 = k_x^2 + k_z^2$. 
With the trial wavefunction $\psi (y) = \psi_{\kappa} e^{\kappa y}$, which is localized at the surface $y=0$,
we obtain the eigenvalue equation
\begin{eqnarray} \label{secEqA}
  \widetilde{H} (k , -i \partial_y )  \psi ( y) = E \psi ( y) ,
\end{eqnarray}
where we have replaced $k_y$ by $- i \partial_y$. Solving the secular equation,
$\mathop{\mathrm{det}} \left[ \widetilde{H} ( k_x, - i \kappa, k_z ) - E \mathbbm{1} \right] = 0$, gives
four solutions for $\kappa (E)$ denoted as $\beta \kappa_{\alpha}(E) $, with $\alpha \in \left\{ 1,2 \right\}$, 
$\beta \in \left\{ +, -\right\}$,  and
\begin{eqnarray} \label{defKap}
\kappa_{\alpha} ( E) 
=
\frac{1}{t} \left[
2 \Delta_{\textrm{t}}^2 - 2 L + k^2 t^2  + (-1)^{\alpha } 2 R 
\right]^{\frac{1}{2} } ,
\end{eqnarray}
where we have introduced the shorthand notation $L = (3 t - \mu ) t$
and $R= \sqrt{ \Delta_{\textrm{t}}^4 - 2 \Delta^2_{\textrm{t}} L  + E^2 t^2 }$. 
For each of the four roots $\kappa_{\alpha} (E)$, the kernel of the secular equation is spanned
by two independent basis vectors, given by
\begin{subequations}
\begin{eqnarray}
\psi_{\alpha \beta 1} 
&=& 
\begin{pmatrix}
\Delta_{\textrm{t}} t k_z \cr
\Delta
_{\textrm{t}} t ( \beta  \kappa_\alpha + k_x ) \cr
0 \cr
E t - L + \frac{t^2}{2} ( k^2 - \kappa_\alpha^2) \cr
\end{pmatrix} ,
\\
\psi_{\alpha \beta 2} 
&=& 
\begin{pmatrix}
\Delta_{\textrm{t}} t ( \beta \kappa_\alpha - k_x ) \cr
\Delta_{\textrm{t}} t k_z \cr
E t - L + \frac{t^2}{2} ( k^2 - \kappa_\alpha^2 ) \cr
0 \cr
\end{pmatrix} .
\end{eqnarray} 
\end{subequations}
With this, we obtain the following ansatz for the surface state wave function
\begin{eqnarray}
\Psi ( {\bf k}
_{\parallel} , y)
=
\sum_{\alpha,\gamma  \in  \{1,2 \} } \sum_{\beta = \pm }
C_{\alpha \beta \gamma} \, \psi_{\alpha \beta \gamma} \, e^{\beta \kappa_{\alpha} y} ,
\quad
\end{eqnarray}
where ${\bf k}_{\parallel} = (k_x, k_z)$ and the coefficients $C_{\alpha \beta \gamma}$  are determined by the boundary conditions 
$\Psi (  {\bf k}_{\parallel} , y=0) =0$ and $\Psi (  {\bf k}_{\parallel} , y \to + \infty ) =0$. The secular equation for the coefficients $C_{\alpha \beta \gamma}$ leads to
the condition
\begin{eqnarray} \label{condMajo}
2 L = \left ( k^2 + \kappa_1 \kappa_2 \right) t^2 ,
\end{eqnarray}
which together with Eq.~\eqref{defKap} yields the dispersion for the surface states
 \begin{eqnarray} \label{HS_disp}
E_{\pm} ( {\bf k}_{\parallel} )
= \pm    \Delta_{\textrm{t} }   k .
\end{eqnarray}
The surface state wave functions at the $\Gamma$ point are found to be
\begin{subequations} \label{eq:SurfaceWavefun}
\begin{eqnarray}
&&
\Psi^{\pm}_{\textrm{MS}} ( {\bf k}_{\parallel} =0 , y ) 
=
\varphi^{\pm}_{\textrm{MS}}
\left[
e^{- \kappa_1 (0) y}  - e^{- \kappa_2 (0) y} 
\right] , \;  \; \;
\\
&&
\varphi^+_{\textrm{MS}}
=
\frac{1}{\sqrt{2}} 
\begin{pmatrix}
1 \cr 0 \cr 1 \cr 0 
\end{pmatrix},
\qquad
\varphi^-_{\textrm{MS}}
=
\frac{1}{\sqrt{2}} 
\begin{pmatrix}
0 \cr 1 \cr 0 \cr 1 
\end{pmatrix} ,
\end{eqnarray}
\end{subequations}
where $\kappa_{\alpha}(0)$ is defined in Eq.~\eqref{defKap}. From  condition~\eqref{condMajo} it follows that zero-energy surface states can exist if $\kappa_1(0)$ and $\kappa_2(0)$ are either both real or complex conjugate partners. In the former case, i.e., when $\Delta^2_{\textrm{t}} > 2L >0$, the wave functions decay monotonically into the bulk with the decay lengths $\kappa^{-1}_{\alpha}(0)$. For $\Delta^2_{\textrm{t}} < 2L$, on the other hand,  $\kappa_{\alpha}(0)$ are complex, which leads to an oscillatory decay of the wave functions with inverse decay length $ \mathrm{Re} [ \kappa_{\alpha}(0) ] =  \Delta_{\textrm{t}} / t $ and oscillation frequencies $\mathrm{Im} [ \kappa_{\alpha}(0) ]$.   
 
An effective low-energy Hamiltonian for the surface states $\Psi^{\pm}_{\textrm{MS}} ( {\bf k}_{\parallel} )$ can be derived by 
projecting $\widetilde{H} ( {\bf k} )$, see Eq.~\eqref{modDefTildO}, onto the subspace $\Psi_{\textrm{MS}} = \left\{ \varphi^+_{\textrm{MS}}, \varphi^-_{\textrm{MS}} \right\}$ formed by the two
surface-state wave functions~\eqref{eq:SurfaceWavefun}. This yields an effective $2 \times 2$ Hamiltonian:
\begin{eqnarray} \label{effHammMajo}
\left\langle \Psi_{\textrm{MS}} \right| \widetilde{H} ( {\bf k} ) \left| \Psi_{\textrm{MS}} \right \rangle
=
\Delta_{\textrm{t}} \left( k_z \sigma_1 - k_x \sigma_3 \right) , \; \; \;  \; \;
\end{eqnarray} 
which has the same dispersion as Eq.~\eqref{HS_disp}. In order to compute the ordinary and spin-resolved FT-STS for the helical Majorana states described by Eq.~\eqref{effHammMajo}, we need to project the impurity potential $V^{\beta}_{\bf q}$, Eq.~\eqref{defHimp}, 
onto the surface-state subspace $\Psi_{\textrm{MS}}$. Using Eq.~\eqref{DefSpinOp}, we find
\begin{eqnarray} \label{impProMS}
\left\langle \Psi_{\textrm{MS}} \right| V^{\beta }_{\bf q}  \left| \Psi_{\textrm{MS}} \right \rangle
= 
\left\{
\begin{array}{c l}
v_{\bf q} \sigma_2 & \; \; \textrm{if $\beta=2$} \cr

0^{\phantom{DD}} & \; \; \textrm{otherwise. } 
\end{array} 
\right.
\end{eqnarray} 
Remarkably, it follows that nonmagnetic scattering is absent, whereas for magnetic impurities only
the out-of-plane spin component  couples to the surface states.
Finally,  to evaluate the FT-STS response, Eq.~\eqref{surfDosA}, we also need to determine
the electronlike parts of the charge and spin operators~\eqref{DefSpinOp} within the surface-state subspace. We have
\begin{eqnarray} \label{SmsOppp}
\left\langle \Psi_{\textrm{MS}} \right| P_{\textrm{e}} S^{\alpha}  \left| \Psi_{\textrm{MS}} \right \rangle
=
\frac{1}{2} \sigma_{\alpha},
\end{eqnarray}
where  $P_{\textrm{e}} = \frac{1}{2} \left( \mathbbm{1}+ \sigma_3  \otimes \sigma_0 \right)$.

\subsection{Arc surface states}  
\label{appendixAb}

Second, we study arc surface states that exist, for example, at the surface of centrosymmetric superconductors with 
point nodes in the BdG excitation spectrum. 
For concreteness, we consider a 
system with a ${\bf d}$-vector ${\bf d}_{\bf k}$ given by Eq.~\eqref{SOCtetragonal}.
Furthermore, neglecting the effects of noncentrosymmetricity, we set the SOC strength $\lambda$ and the spin-singlet component $\Delta_{\textrm{s}}$ of the superconducting gap to zero. 
Within a small-momentum expansion near $k_x=k_y=0$, the superconductor is described by
\begin{eqnarray} \label{linearArcHam}
\widetilde{H} ( \bm{k} )
=
\begin{pmatrix}
\tilde{\varepsilon}_{\bf k} \sigma_0 &  
- \Delta_{\textrm{t}} ( k_y \sigma_3 + i k_x \sigma_0 )  \cr
- \Delta_{\textrm{t}} ( k
_y \sigma_3 - i k_x \sigma_0 ) & - \tilde{\varepsilon}_{\bf k} \sigma_0 \cr   
\end{pmatrix} ,
\nonumber\\
\end{eqnarray}
where $\tilde{\varepsilon}_{\bf k} =2 t + t  \cos k_z  - \mu - \frac{t}{2} ( k_x^2 + k_y^2 )$.  
As before, the ansatz for the surface-bound-state wave function is taken to be $\Psi (y) = \Psi_{\kappa} e^{\kappa y}$, with the inverse decay length $\kappa$. From the secular equation, we obtain four solutions for $\kappa(E)$ given by $\pm \kappa_\alpha (E) $:
\begin{eqnarray} \label{kappaArc}
\kappa_{\alpha} (E)
=
\frac{1}{t}
\left[
2 \Delta_{\textrm{t}}^2 - 2 L+   k_x^2 t^2
+ (-1)^{\alpha} 2 R
\right]^{\frac{1}{2}} ,
\end{eqnarray}
with $\alpha \in \left\{ 1, 2 \right\}$, $R= \sqrt{ \Delta_{\textrm{t}}^4 - 2 \Delta_{\textrm{t}}^2 L+ E^2 t^2}$, and $L= \left( 2 t + t   \cos k_z  - \mu \right) t$.
Repeating similar steps as above, we obtain for the surface-state trial wave function
\begin{eqnarray}
\Psi ( {\bf k}_{\parallel} , y)
=
\sum_{\alpha,\gamma  \in  \{1,2 \} } \sum_{\beta = \pm }
C_{\alpha \beta \gamma} \, \psi_{\alpha \beta \gamma} \, e^{\beta \kappa_{\alpha} y} ,
\quad
\end{eqnarray}
with ${\bf k}_{\parallel} = (k_x, k_z)$ and the two independent vectors
\begin{subequations}
\begin{eqnarray}
\psi_{\alpha \beta 1}
&=&
\begin{pmatrix}
0 \cr
- i \Delta_{\textrm{t}} t \left( \beta \kappa_{\alpha} + k_x \right) \cr
0 \cr
E t - L + \frac{t^2}{2} \left( k_x^2 - \kappa_{\alpha}^2 \right) \cr
\end{pmatrix} ,
\\
\psi_{\alpha \beta 2}
&=&
\begin{pmatrix}
+ i \Delta_{\textrm{t}} t \left( \beta \kappa_{\alpha} - k_x \right) \cr
0 \cr
E t - L + \frac{t^2}{2} \left( k_x^2 - \kappa_{\alpha}^2 \right) \cr
0
\end{pmatrix} .
\end{eqnarray}
\end{subequations}
A surface state occurs if the coefficients $C_{\alpha \beta \gamma}$ can be chosen such that the wave function $\Psi ( {\bf k}_{\parallel} , y)$
satisfies the boundary conditions $\Psi ( {\bf k}_{\parallel} , y=0) =0$ and $\Psi ( {\bf k}_{\parallel}  , y \to + \infty ) =0$.
After some algebra, this leads to the existence condition for the surface states:
\begin{eqnarray} \label{condArc}
2 L = ( k_x^2 + \kappa_1 \kappa_2 ) t^2  .
\end{eqnarray}
Combining Eqs.~\eqref{kappaArc} and \eqref{condArc} gives the dispersion  $E_{\pm} ( {\bf k}_{\parallel} )= \pm    \Delta_{\textrm{t} }   k_x $.
For the surface-state wave functions at $k_x=0$, we find 
\begin{subequations} \label{zeroEWFarc}
\begin{eqnarray}
&&
\Psi^{\pm}_{\textrm{AS}} ( k_{x} =0, k_z , y ) 
=
\varphi^{\pm}_{\textrm{AS}}
\left[
e^{- \kappa_1 ( 0 ) y}  - e^{- \kappa_2 ( 0 ) y} 
\right] , \;  \; \; \; \; 
\\
&&
\varphi^{+}_{\textrm{AS}}
=
\frac{1}{\sqrt{2}}
\begin{pmatrix}
-i \cr
0 \cr
+ 1 \cr
0
\end{pmatrix},
\qquad
\varphi^{-}_{\textrm{AS}}
=
\frac{1}{\sqrt{2}}
\begin{pmatrix}
0 \cr
+ i \cr
0 \cr
+ 1
\end{pmatrix} ,
\end{eqnarray} 
\end{subequations}
where $\kappa_{\alpha} (0)$ is given by Eq.~\eqref{kappaArc}. For $\Delta_{\textrm{t}}^2 > 2 L > 0$, i.e., for
$ \arccos \left[ ( \Delta_{\textrm{t}}^2 / (2 t) - 2 t + \mu ) / t  \right] <  \left| k_z \right| < \arccos \left[ ( \mu - 2 t ) / t  \right]$,
the zero-energy wave functions \eqref{zeroEWFarc} decay exponentially and monotonically into the bulk, whereas for $2L > \Delta^2_{\textrm{t}}$ the exponential wave function decay is modulated by
periodic oscillations with frequencies $\mathrm{Im} \left[ \kappa_{\alpha} (0) \right]$.

Projecting $\widetilde{H} ( {\bf k} )$, see Eq.~\eqref{linearArcHam},  onto the subspace
$\Psi_{\textrm{AS}} = \left\{ \varphi^-_{\textrm{AS}} , \varphi^+_{\textrm{AS}} \right\}$ yields a low-energy effective
Hamiltonian for the arc surface states: 
\begin{eqnarray} \label{projArcH}
\left\langle \Psi_{\textrm{AS}} \right| \widetilde{H} ( {\bf k} ) \left| \Psi_{\textrm{AS}} \right \rangle
=
- \Delta_{\textrm{t}} k_x \sigma_3 . 
\end{eqnarray}
The disorder potential $V^{\beta}_{\bf q}$, see Eq.~\eqref{defHimp}, within the surface-state subspace $\Psi_{\textrm{AS}}$ reads
\begin{eqnarray}
\left\langle \Psi_{\textrm{AS}} \right|  V_{\bf q}^{\beta}   \left| \Psi_{\textrm{AS}} \right \rangle
=
\left\{
\begin{array}{c l}
- v_{\bf q} \sigma_1 & \; \; \textrm{if $\beta=1$, } \cr
0^{\phantom{DD}} & \; \; \textrm{otherwise, } 
\end{array} 
\right.
\end{eqnarray}
whereas the projected  electronlike parts of the charge and spin operators~\eqref{DefSpinOp} are given by
\begin{eqnarray} \label{projSaCopAS}
\left\langle \Psi_{\textrm{AS}} \right| P_{\textrm{e}} S^{\alpha}  \left| \Psi_{\textrm{AS}} \right \rangle
=
\frac{1}{2} (-1)^{\alpha} \sigma_{\alpha} ,
\end{eqnarray}
where  $P_{\textrm{e}} = \frac{1}{2} \left( \mathbbm{1}+ \sigma_3  \otimes \sigma_0 \right)$.

\section{Symmetries of  QPI patterns}
\label{appendixC}

\subsection{Time-reversal symmetry}

Time-reversal symmetry acts on the single-particle Green's function 
$G^{(0)} ( {\bf k}_{\parallel} ) = \left[ E + i \eta - H  ({\bf k}_{\parallel} ) \right]^{-1}$ % 
as
\begin{eqnarray} \label{TRSc1}
U_T  \big[ G^{(0)} (-  {\bf k}_{\parallel} ) \big]^{\textrm{T}}   U^{\dag}_T 
=
G^{(0)} ( {\bf k}_{\parallel} ) ,
\end{eqnarray}
where $U_T = \sigma_0 \otimes i \sigma_2$. Inserting relation~\eqref{TRSc1} into the definition of $\delta \rho_{\textrm{s}}$, see Eq.~\eqref{FTSTSresponse}, yields
\begin{align}
\delta \rho_{\textrm{s}} 
\left( U^{\dag}_T \big[ S^{\alpha} \big]^{\textrm{T}} U_T, U^{\dag}_T \big[ V^{\beta}  \big]^{\textrm{T}} U_T ; {\bf  q}_{\parallel} \right)
=
\delta \rho_{\textrm{s}} ( S^{\alpha}, V^{\beta}; {\bf q}_{\parallel} ),
\end{align}
where we have explicitly written out the dependence of $\delta \rho_{\textrm{s}}$ on the spin operations $S^{\alpha}$ and
the impurity potential $V^{\beta}$.
Since magnetic impurity potentials are odd under time-reversal symmetry, whereas $S^0$ is even, it follows that, within the Born approximation, $\delta \rho^{0 \beta}_{\textrm{s}} =0$, for $\beta \in \left\{ 1,2,3 \right\}$. Similarly, we have $\delta \rho^{\alpha 0}_{\textrm{s}} =0$, for $\alpha \in \left\{ 1,2,3 \right\}$.

\subsection{Point-group symmetries}

As in the main text, we consider  FT-STS patterns on a surface that is perpendicular to the $y$ axis. 
Denoting those symmetry operations of the crystallographic point group that leave the (010) plane invariant by $R$,
we find that  $\delta \rho^{\ }_{\textrm{s}}$ transforms under $R$ as~\cite{Schnyder2012,samokhin09}
\begin{equation} \label{symPropQPI}
\delta \rho_{\textrm{s}} 
\left(  U^{\dag}_{\widetilde{R}} S^{\alpha} U^{\phantom{\dag}}_{\widetilde{R}} , U^{\dag}_{\widetilde{R}}  V^{\beta} U^{\phantom{\dag}}_{\widetilde{R}} ;  R^{-1} {\bf q}_{\parallel}   \right) 
=
\delta \rho_{\textrm{s}} \left( S^{\alpha}, V^{\beta} ; {\bf q}_{\parallel} \right) ,
\end{equation}
with $\widetilde{R} = \det (R) R$, $U_{\widetilde{R}} = \mathop{\textrm{diag}}  ( u^{\phantom{*}}_{\widetilde{R}}, u^{\ast}_{\widetilde{R}}  )$,
and $u_{\widetilde{R}} = \exp \left[ - i ( \Theta /2) \hat{\bf n} \cdot \bm{\sigma} \right]$, where $\Theta$ is the angle of rotation of $R$
and $ \hat{\bf n}$ denotes the unit vector along the rotation axis.

\subsubsection{Cubic point group $O$} 

For the cubic point-group $O$ with ${\bf l}_{\bf k}$ given by Eq.~\eqref{SOCcubic}, 
Hamiltonian~\eqref{modelDef} is invariant under $\pi/2$ rotations about the $y$ axis. Making use of Eq.~\eqref{symPropQPI} with
$R = R_{\hat{{\bf y}}}$, where
\begin{align}
R_{\hat{{\bf y}}} 
=
\begin{pmatrix}
0 & 0 & +1 \cr
0 & +1 & 0 \cr
-1 & 0 & 0 \cr
\end{pmatrix},
\end{align}
 yields
\begin{align}
&&
\delta \rho_{\textrm{s}}^{11}({\bf q}_{\parallel} ) = +\delta \rho_{\textrm{s}}^{33}(R^{-1}_{\hat{{\bf y}}}  {\bf q}_{\parallel} ),
\;
\delta \rho_{\textrm{s}}^{12}({\bf q}_{\parallel} )=- \delta \rho_{\textrm{s}}^{32}(R^{-1}_{\hat{{\bf y}}}  {\bf q}_{\parallel} ),
\quad \;  \; \;
\nonumber\\
&&
\delta \rho_{\textrm{s}}^{21}({\bf q}_{\parallel} )=- \delta \rho_{\textrm{s}}^{23}(R^{-1}_{\hat{{\bf y}}}  {\bf q}_{\parallel} ),
\;
\delta \rho_{\textrm{s}}^{13}({\bf q}_{\parallel} )=- \delta \rho_{\textrm{s}}^{31}(R^{-1}_{\hat{{\bf y}}}  {\bf q}_{\parallel} ) .
\quad \;  \; \;
\end{align}
Furthermore, we find that the QPI patterns  $ \delta \rho_{\textrm{s}}^{00}({\bf q}_{\parallel} )$ and $ \delta \rho_{\textrm{s}}^{22}({\bf q}_{\parallel} )$
are invariant under $\pi / 2$ rotations, i.e., $\rho_{\textrm{s}}^{\alpha \alpha}({\bf q}_{\parallel} )=\rho_{\textrm{s}}^{\alpha \alpha}(R^{-1}_{\hat{{\bf y}}}   {\bf q}_{\parallel} )$,
for $\alpha \in \left\{ 0, 2 \right\}$.

\subsubsection{Monoclinic point group $C_{2}$}  \label{appendixCzwei}

In the case of the monoclinic point-group $C_{2}$ with ${\bf l}_{\bf k}$ given by Eq.~\eqref{SOCmonoclinic}, we find 
that the (010) and (100) faces are equivalent, i.e., the zero-energy states appearing on these two surfaces are identical.
Hence we consider $\pi $ rotations about the $\frac{1}{ \sqrt{2} } ( \hat{{\bf x}} + \hat{{\bf y}} )$ axis with
\begin{align}
R_{\hat{{\bf x}} +  \hat{{\bf y}}  }
=
\begin{pmatrix}
0 & +1 & 0 \cr
+1 & 0 & 0 \cr
0 & 0 & -1 \cr
\end{pmatrix}
\end{align}
 to obtain the following symmetry relations for $\delta \rho_{\textrm{s}}$:
\begin{eqnarray}
\delta \rho_{\textrm{s}}^{11} ( {\bf q}_{\parallel} ) =  + \delta \rho_{\textrm{s}}^{22} ( \widetilde{{\bf q}}_{\parallel}  ) ,
\;
\delta \rho_{\textrm{s}}^{12} ( {\bf q}_{\parallel} ) =  + \delta \rho_{\textrm{s}}^{21} ( \widetilde{{\bf q}}_{\parallel}  ) ,
\nonumber\\
\delta \rho_{\textrm{s}}^{13} ( {\bf q}_{\parallel} ) =  - \delta \rho_{\textrm{s}}^{23} ( \widetilde{{\bf q}}_{\parallel}  ) ,
\;
\delta \rho_{\textrm{s}}^{31} ( {\bf q}_{\parallel} ) =  - \delta \rho_{\textrm{s}}^{32} ( \widetilde{{\bf q}}_{\parallel}  ) ,
\end{eqnarray}
where  $\widetilde{{\bf q}}_{\parallel} = ( q_x, - q_z)$.

\bibliography{paper}

 \end{document}